\documentclass[%
 reprint,
 amsmath,amssymb,
 aps,
]{revtex4-2}
\usepackage{float}
\usepackage{comment} 
\usepackage{graphicx}
\usepackage{dcolumn}
\usepackage{bm}
\usepackage[pdftex, pdftitle={Article}, pdfauthor={Author}]{hyperref} 
\usepackage{hyperref}
\usepackage[table]{xcolor}

\begin{document}
\preprint{APS/123-QED}
\title{Controlling encirclement of an exceptional point \\ using coupled spintronic nano-oscillators }

\author{K. Ho$^{1}$,  S. Perna$^{2}$, S. Wittrock$^{3}$, S. Tsunegi$^{4}$, \\
H. Kubota$^{4}$, S. Yuasa$^{4}$, P. Bortolotti$^{1}$, M. d'Aquino$^{2}$, C. Serpico$^{2}$, V. Cros$^{1}$, R. Lebrun$^{1}$ \\ \normalsize 
\textit{$^{1}$Laboratoire Albert Fert,
CNRS, Thales, Université Paris-Saclay, Palaiseau, 91767, France} \\ \normalsize 
\textit{$^{2}$Department of Electrical Engineering and ICT, University of Naples Federico II, Naples, 80125, Italy} \\ \normalsize 
\textit{$^{3}$Helmholtz-Zentrum Berlin fur Materialien und Energie GmbH, Hahn-Meitner-Platz 1, Berlin, 14109, Germany}  \\ \normalsize 
\textit{$^{4}$National Institute of Advanced Industrial Science and Technology (AIST), Tsukuba, Ibaraki 305-8568, Japan} \\ \normalsize 
}

\date{\today}
\begin{abstract} Exceptional points (EPs), branch singularities parameter space of non-Hermitian eigenvalue manifolds, display unique topological phenomena linked to eigenvalue and eigenvector switching: the parameter space states are highly sensitive to the system's parameter changes. Therefore, we suggest investigating the parameter space in the presence of an EP by experimentally accessing and exploiting the topological nature of the coupled system around an EP. We demonstrate control over exceptional points in coupled vortex spin-transfer torque oscillators by adjusting the system's damping through the spin-transfer torque effect and their relative phase. This approach allows for precise manipulation of the coupling behavior in the vicinity of an exceptional point.   We report the presence of both level attraction/repulsion by adjusting the system's parameters. Moreover, we evidence the topological nature of the EP by dynamically encircling it in the phase-current parameter space, leading to a switch of the eigenstates. Our study introduces a new method for exploring non-Hermitian physics in spintronic systems at room temperature.

\end{abstract}
\maketitle


Exceptional points (EPs), defined as degeneracies of non-Hermitian systems at which both eigenvalues and eigenvectors coalesce \cite{kato2013perturbation,heiss_physics_2012}, can emerge from dissipative open systems that exhibit parity-time (PT) symmetry. EPs can thus arise in different types of physical systems such as those studied in photonics \cite{miri2019exceptional, ozdemir_paritytime_2019,de_carlo_non-hermitian_2022}, spintronics \cite{ Tserkovnyak2020-hh, flebus_non-hermitian_2020,deng_exceptional_2022} and cavity magnonics \cite{Liu2019-ba, wang_dissipative_2020, harder_coherent_2021}. A theoretical understanding of EPs and thoroughly exploring the parameter space surrounding these singularities are crucial for controlling  emerging phenomena with potential innovative applications, for instance, enhanced sensitivity near the EPs \cite{ hodaei_enhanced_2017, wiersig_review_2020} and non-reciprocal energy conversion  \cite{xu_topological_2016, li_exceptional_2023}.

In non-Hermitian systems, complex energy levels can either attract or repeal each other depending on the nature of the coupling,  namely conservative vs. dissipative \cite{Bernier2018,zhang_dissipative_2022}. Indeed, controlling the effective dissipation in the system allows the engineering of dissipative coupling, which can enable reaching an EP in a non-Hermitian system \cite{ heiss_phases_1999, Lee2009-fq, Cao2015-wr}. In the region of conservative coupling, level repulsion can be observed, with a typical avoided crossing pattern \cite{boventer_steering_2019}. Conversely, level attraction and coalescence are signatures of systems' dissipative nature \cite{wang_dissipative_2020}. 

The transition from level repulsion to level attraction is typically marked by the appearance of EPs \cite{Maznev2018}. Due to the Riemann surface topology of the parameter space and the intersecting eigenvalue manifolds on the branching cut departing from an EP, the encircling of an EP results in a path through the branching cut and eventually in a switching of the system's eigenstate \cite{heiss_physics_2012, Hassan2017-zk}. Such adiabatic dynamic encircling, which explores the topological nature of the parameter space around an EP,  has been demonstrated in a few physical systems such as microwave cavities \cite{Persson2000, dembowski_observation_2003} or waveguides \cite{doppler_dynamically_2016, Li2020-sk}. 

However, identifying and controlling the complex topology of parameter spaces around EPs remains challenging and has largely hindered their integration into CMOS-compatible electronic devices. Spintronic nano-oscillators emerge as a promising candidate for overcoming these challenges, paving the way for the manipulation and exploration of EPs in nanoscale devices. Over the past decade, this class of oscillators has garnered significant attention for applications in wireless communication \cite{sharma_electrically_2021}, random number generation \cite{PhanPhysRevApplied.21.034063}, and  neuromorphic computing \cite{Locatelli2014-bj, Romera2018, Grollier2020NeuromorphicS}.
Recent studies have revealed that the phenomenon of oscillation death in strongly coupled STVOs can be attributed to the presence of an exceptional point, as described within the framework of Thiele formalism \cite{wittrock_non-hermiticity_2024}. Exploiting the intricate topology of the parameter space around the EP, however, demands precise control over the trajectory topology
within this space. This level of control is achieved in the present work
by using a system comprising amplifiers and a phase shifter between two coupled STVOs.

In this Letter, we experimentally demonstrate the dynamical encircling of an exceptional point in two coupled vortex-based spintronic oscillators. 
First, we provide experimental evidence of the transition in coupling nature, from conservative to dissipative, achieved by tuning the effective phase shift of their coupling. By defining an appropriate closed loop in the phase-current parameter space, we then observe mode exchange between the two Riemann sheets, manifested as a frequency shift of the coupled oscillating regimes. Finally, we provide experimental evidence that the system can recover its initial state after a second similar loop in the parameter space. 

Our methodology enables the exploration of non-Hermitian physics using a lumped-element approach, significantly simplifying the study of  EPs in nanoscale CMOS-compatible circuits at room temperature.

\begin{figure}[!htb]
    \centering
\includegraphics[width=8.5cm]{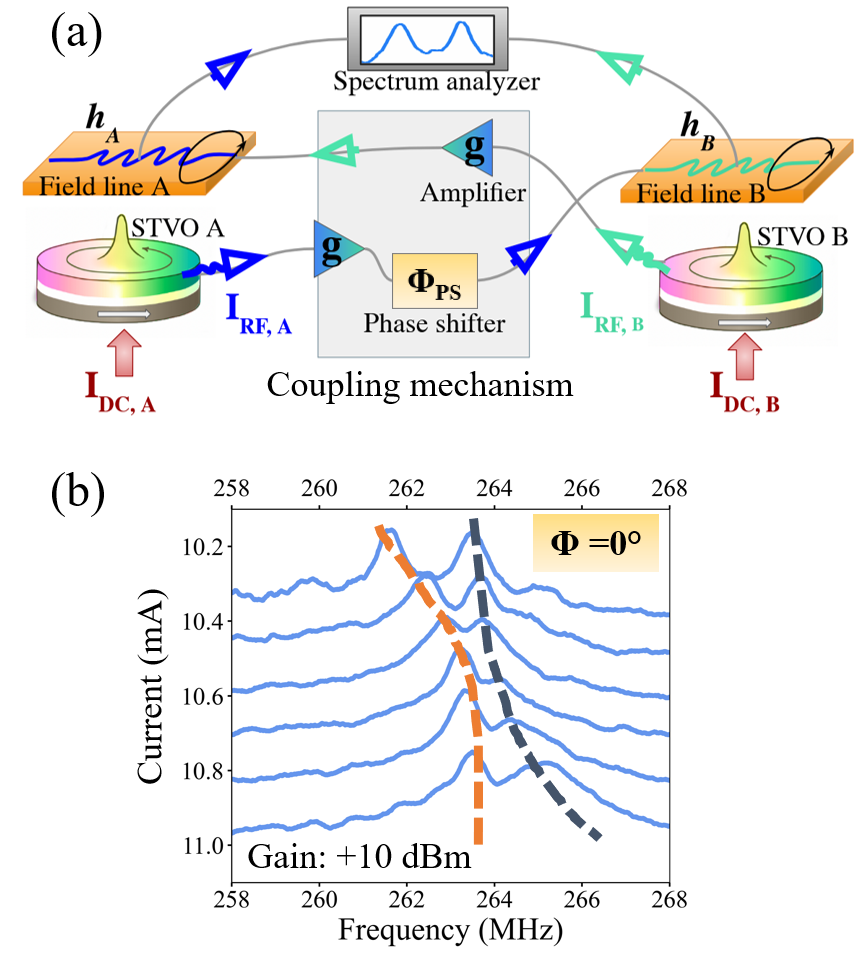}
\caption{\label{fig:fig1} (a) Schematic of the coupled oscillators. Each STVO auto-oscillates under the application of a DC current $I_{DC}$.  One STVO is supplied with constant current while the other is swept across an increasing current range. The RF signal $I_{RF}$ from each STVO is conveyed into the other's field line to produce an RF field $h$. The coupled signal is measured with a spectrum analyzer. A variable amplifier, set at +10 dBm, controls the coupling amplitude $g$; a phase shifter $\Phi_{PS}$ tunes the dephasing $\Phi$. (b)  Typical experimental spectra (blue) of level repulsion at varying DC currents: a gap separates the hybridized modes (orange and blue), which bend away.  The intrinsic phase of the coupled system is equal to $\Phi_g = -10$°. }
\end{figure}

The experimental setup 
is illustrated in Fig.\ref{fig:fig1}(a). This configuration allowed us to devise a strategy for controlling the presence of EPs in coupled STVOs driven by their self-generated radiofrequency (rf) currents. The gain and loss mechanisms for each oscillator are governed both by the natural magnetization damping and the spin-transfer torques (STT). To achieve the critical condition required for reaching an EP, we vary the injected DC current in one oscillator while keeping it fixed in the other.
The strength and phase of the coupling between the two spintronic oscillators can be regulated using a variable gain amplifier and variable phase inserted in the RF current lines (see Fig.~\ref{fig:fig1}(a)).

The dynamics of the coupled oscillators, labeled with $l$=(A, B), can be modeled within the Thiele formalism by using gyration of the vortex cores from the disks' centers as dynamical variables.  In this respect, we introduce the complex variables $z_l = {X_l}/{R} + i {Y_l}/{R}$,  where $X_l, Y_l$ are Cartesian coordinates of the vortex cores displacements, and $R$ is the STVO radius \cite{Perna2024-dg, wittrock_non-hermiticity_2024}. When small oscillations of the vortex cores are considered, the governing equations can be linearized around their equilibrium positions, and this leads to the coupled system:
\begin{align}
\label{eq:one}
\frac{d}{dt}
\begin{bmatrix}
z_A\\
z_B
\end{bmatrix}	= i
\begin{bmatrix}
\omega_A -i\gamma_A & g e^{i\Phi_g}\\
g e^{i(\Phi-\Phi_g}) & \omega_B -i\gamma_B 
\end{bmatrix}	
\begin{bmatrix}
z_A\\
z_B
\end{bmatrix}
\end{align}
where $\omega_l=\omega_l(I_l)$
are the angular frequencies of the two oscillators
$\gamma_l =  C_l I_l  - d_l\omega_{l}(I_{l}) $
are the gain/loss coefficients, 
where $d_l$ are  damping  coefficients and 
$C_{l}$ are STT efficiency factors.
The parameter $g$, which is real and positive,
characterizes the coupling
strength while $\Phi_g$
is a bias in the phase
shift. 
The real-valued phase
$ \Phi \in [0,2\pi] $ controls  the nature of the coupling. Specifically, it is possible to go from conservative coupling  for $\Phi=0$ (level repulsion) 
to purely dissipative coupling for $\Phi=\pi$ (level attraction).

\begin{figure*}[!htb]
    \centering
\includegraphics[width=16.5cm]{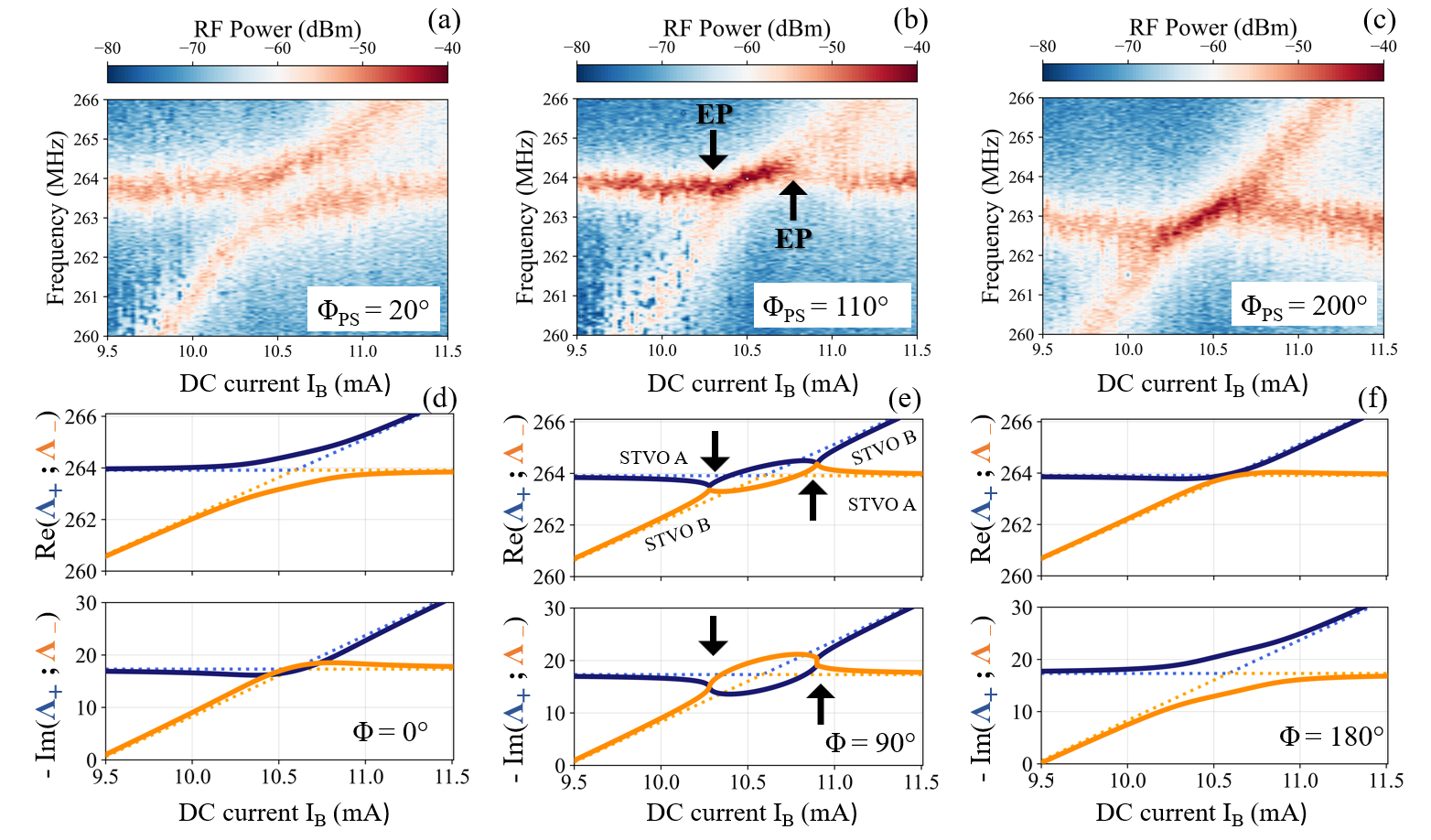}
\caption{\label{fig:fig2}Experimental maps of  frequencies vs injected DC current  from level repulsion at $\Phi_{PS} = 20$° (a), to the transition marked by two EPs at $\Phi_{PS} = 110$° (b) and level attraction at $\Phi_{PS} = 200$° (c). Corresponding theoretical real (top) and imaginary (bottom) parts of the eigenvalues $\Lambda_{\pm}(I_B, \Phi)$ evolving with the DC current (d-f), representing the frequencies and stability from $\Phi = 0$° to $\Phi = 180$°, with $\Phi = \Phi_{PS} + 2\Phi_{g}$. $\Phi_{g} = -10$° encompasses the contributions of electrical length and spin-transfer torques. Solid lines indicate the coupled modes; dashed lines are the uncoupled modes. Two EPs bound the region indicated by arrows, where the imaginary and real parts of the eigenvalues coalesce. The gain is set at +10 dBm. }
\end{figure*}

Note that the experimentally controlled phase shift between the two oscillators is given by $\Phi_{PS} = \Phi-2\Phi_g$, where $\Phi_g$ is the intrinsic phase of the coupled system. The phase $\Phi_g$ includes the contributions of electrical length between the oscillators and the conservative and dissipative effects of the damping-like and field-like spin-transfer torques \cite{Lebrun2017, Hem2019}. In our experiments,  $\Phi_{g}$ is equal to -10°.  For example, for $\Phi_{PS}=20$°, shown in Fig.~\ref{fig:fig1}(b), the modes of the two STVOs repel each other, corresponding to the case of conservative coupling ($\Phi=0$). In Fig.~\ref{fig:fig2}(a-c), discussed further later, experimental emission spectra of the coupled system for three values of $\Phi_{PS}$ show the transition from level repulsion to level attraction  over a 
$\Phi$-cycle. 



\smallbreak

 In order to detect the presence of exceptional points and to determine the conditions for their appearance, we correlate the experimental observation of the oscillators’ frequencies with the real and imaginary parts of the eigenvalues $(\Lambda_{+},\Lambda_{-})$
 of the matrix and the right-hand-side of eq.\eqref{eq:one}. 
 The real parts of the eigenvalues correspond to the mode frequencies, while their imaginary parts control the stability.
 As the parameters are varied, a Hopf bifurcation occurs when the imaginary parts of the eigenvalues transition from negative to positive values. This bifurcation 
leads to the emergence of self-oscillating regimes. For parameter values 
close to the Hopf bifurcation point, where no additional bifurcations occur, linear analysis can be employed to estimate the frequency of these self-oscillating regimes.
In interpreting the experimental results, this concept is applied to link the  real parts of the eigenvalues to the spectral peaks observed in 
the self-oscillating regimes \cite{Wiggins}.

To proceed with our discussion, the eigenvalues are represented in the complex plane as functions of the control parameters $(I_A, I_B, g, \Phi_{PS})$. Their behaviors are to the coupled dynamics. 
The eigenvalues  are given by: 
\begin{eqnarray}
\Lambda_{\pm} = \bar{\omega} - i \bar{\gamma} 
\pm \sqrt{ \left(  \widetilde{\omega}-i\widetilde{\gamma} \right)^2 + 
g^2e^{i\Phi} }
\label{eq:eigenvalue} 
\end{eqnarray}
where $ \bar \gamma = ({\gamma_A + \gamma_B})/{2}$, $\bar \omega = ({\omega_A + \omega_B})/{2}$ and $\widetilde \gamma = ({\gamma_A - \gamma_B})/{2}$, $\widetilde \omega = ({\omega_A - \omega_B})/{2}$. 
Eq.(\ref{eq:eigenvalue}) clearly shows that the eigenvalues do not depend on $\Phi_g$. The condition to observe EPs is the following:
\begin{eqnarray}
g^2e^{i\Phi} + \left(\widetilde \omega-i\widetilde \gamma\right)^2 =0
\label{eq:condEP}
\end{eqnarray}
at which the eigenvalues' real and imaginary parts coincide, with the coalescence of both eigenvalues and eigenvectors. The condition for the presence of EPs can hence be experimentally reached by fine-tuning 
the system's parameters. 
In that respect, our ability to precisely shift the coupling relative phase $\Phi_{PS}$ is a critical tool for controlling the coupling nature and how the modes tend to repel or attract each other through the presence of EPs.

In Fig.~\ref{fig:fig2}, we describe how, by controlling $\Phi_{PS}$, to reach the conditions to reach an EP in the transition from level repulsion to level attraction. For a phase fixed at $\Phi_{PS}=20$° ($\Phi=0$), see Fig.~\ref{fig:fig2}(a) (experiments) and Fig.~\ref{fig:fig2}(d) (theory), we find that the STVO frequencies bend away from each other due to a conservative contribution that largely dominates. In this case, the real part of the eigenvalues repel while the imaginary parts attract. The anticrossing behavior results from the avoidance of degeneracies. As a result, no EPs are observed in the first case. From this figure, we identify the relevant parameters in our oscillators model that are kept fixed in the analysis of the other cases.  By varying the phase shifter by $\frac{\pi}{2}$, leading to $\Phi_{PS} = 110$° ($\Phi=\frac{\pi}{2}$), two EPs are reached in the range $I_{B}$=[10.3, 10.8] mA as shown in Fig.~\ref{fig:fig2}(b) and (e). In this case, we have mixed contribution of conservative and dissipative coupling. The attraction of the energy levels can be seen precisely at the EPs indicated by arrows, as in Fig.~\ref{fig:fig2}(e) where $\text{Re}(\Lambda_+) = \text{Re}(\Lambda_-) $ and  $\text{Im}(\Lambda_+) = \text{Im}(\Lambda_-) $ coalesce. For $\Phi_{PS}=200$°  ($\Phi=\pi$), we observe in Fig.~\ref{fig:fig2}(c) that the modes frequencies merge in a range of injected current of about 1 mA. This is the typical situation of level attraction with
merging of the real parts and anticrossing of the imaginary parts as represented in Fig.~\ref{fig:fig2}(f). This corresponds to a purely dissipative coupling. 
It is important to highlight that the single-peak response observed through dissipative coupling originates from level attraction in the spectrum of the linearized equations. This mechanism is fundamentally different from the nonlinear process that produces singly peaked spectra arising from the synchronization of two (or more) spintronic oscillators, as described in \cite{Lebrun2017}.

It can be noted that level repulsion and attraction are consistently shifted by $\pi$ and are separated by regions when two EPs are reached. Moreover, the full system exhibits $2\pi$-periodicity, allowing for the possibility of observing these different cases twice over an entire $\Phi$-cycle.

\begin{figure*}[htb!]
    \centering
\includegraphics[width=16.5cm]{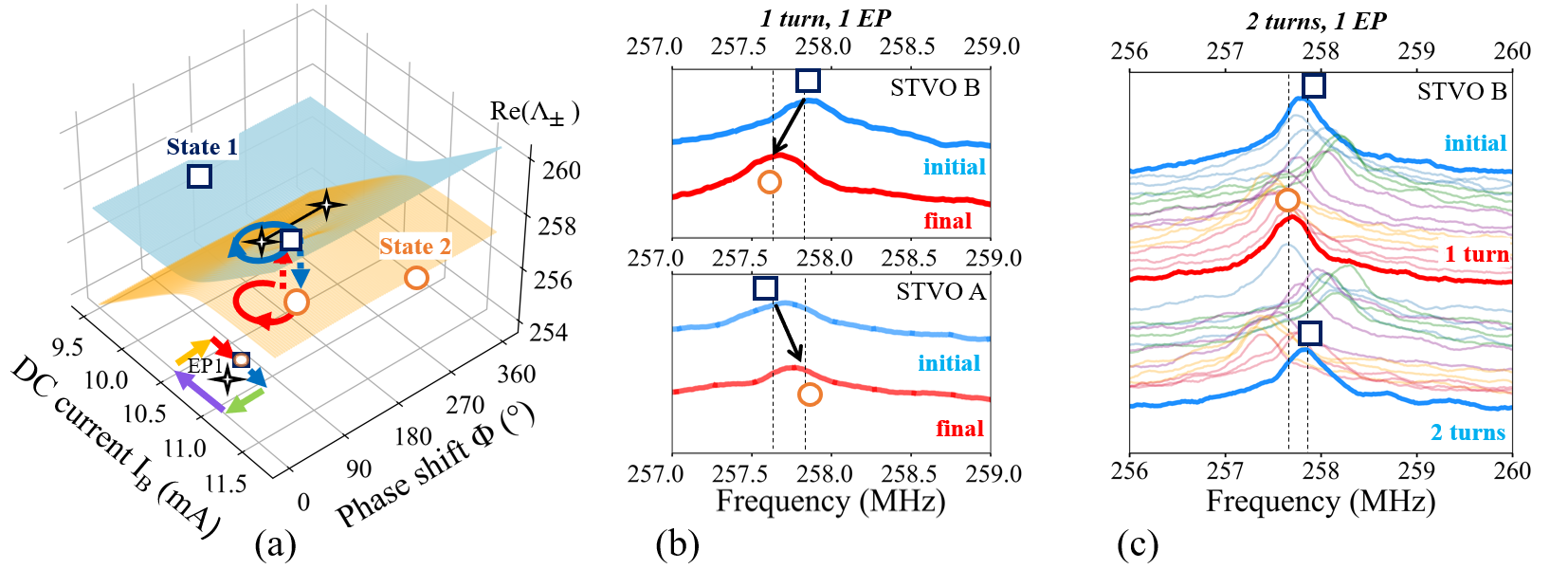}
\caption{\label{fig:fig3}Illustration of the eigenstate exchange after encircling an EP in the parameter space. Riemann sheet structure of the eigenvalues real part in the parameter space $(I_B, \Phi)$,  trajectories  (blue and red lines) to encircle an EP (star) (a). Encircling an EP once results in an eigenstate switch from a higher frequency (square) to a lower frequency  (circle) state separated by a frequency shift $\delta \omega$ for STVO $B$ and vice-versa for STVO $A$, evidenced by independent spectra at initial and final steps of encirclement protocol (b).   Corresponding experimental independent spectra at each step of the closed loop (projected arrows in (a)) for encircling an EP twice (c). The initial state is retrieved if the EP is encircled twice. Dashed lines are eye guides. }
\end{figure*}

Next, we aim at exploring the topological properties of the parameter space in the presence of EPs. This will bring further confirmation of the theoretical model which predicts the appearance of EPs. Indeed, the complex root eigenvalue structure of the non-Hermitian system leads to a non-trivial topology in the parameter space. Thus, the system is very sensitive to whether the trajectory encircles an EP or not, as shown in other physical systems \cite{harder_topological_2017, Zhong2018-gc, Wingenbach2024-re}. The complex eigenvalues $\Lambda_{\pm}(I_B,\Phi)$ expressed in Eq. (\ref{eq:eigenvalue}) are represented by a multivalued  function, with a positive and a negative root sheet, connected by a discontinuous branch cut (black line in Fig.~\ref{fig:fig3}(a)), whose two extremities are the two EPs. A necessary consequence is that completing an entire loop around one EP shall result in a transition from one Riemann sheet to the other.  The initial state (blue line in top panel of Fig.~\ref{fig:fig3}(b)) is precisely recovered only after completing 2 turns around the EP (blue line in Fig.~\ref{fig:fig3}(c)).


An essential result of this study is the successful completion of a full parametric closed loop in the parameters' space around an EP, leading to an exchange of eigenstates. To this end, we establish a closed loop by adjusting two parameters: one for coupling, $\Phi$ and one for gain/loss, $I_B$. 
The five-step closed loop in the parameter space, depicted in Fig.~\ref{fig:fig3}(a) by projected arrows, starts from an initial point ($I_{B}^{i}, \Phi^{i}$) in Cartesian coordinates. The process begins by increasing the DC current to ($I_{B}^{i} + \delta I / 2, \Phi^{i}$), then tuning the phase shift to ($I_{B}^{i} + \delta I / 2, \Phi^{i} - \delta \Phi$). The loop continues by adjusting to ($I_{B}^{i} - \delta I / 2, \Phi^{i} - \delta \Phi$), tuning the phase back to ($I_{B}^{i} - \delta I / 2, \Phi^{i}$), and finally restoring the DC current to compare the final state to the initial one at ($I_{B}^{i}, \Phi^{i}$).

In Fig.~\ref{fig:fig3}(b), we present the experimental spectra, recorded independently using one spectrum analyzer for respectively STVO $A$ and $B$ (bottom and top panels). The measurements correspond to the initial (square) and final (circle) steps of the parametric closed loop during one turn around one of the two EPs. For STVO $B$, we observe a lower frequency for the final state (circle) than the initial state (square) after a complete loop. These two states are separated by $\delta \omega$ = 0.4 MHz, evidencing a frequency shift (see dashed lines). The Riemann sheet structure can explain this frequency shift, as performing one closed loop around an EP means crossing the branch cut (black line) and essentially changing the sheet. In the experiment presented in Fig.~\ref{fig:fig3}(b), STVO $B$ starts from the upper state (square) and ends on the lower state (circle) on the bottom sheet with a lower frequency. The inverse is observed for STVO $A$, demonstrating the state exchange after one encirclement protocol. We then repeat this encircling procedure a second time and observe that the oscillator (STVO $B$ in Fig.~\ref{fig:fig3}(c)) returns as expected to its initial state. The detailed steps of the closed loop for one and two turns of the encirclement process, which correspond to the states associated with the trajectories on the Riemann sheets are shown in Fig.~\ref{fig:fig3}(c).


The experimental findings shown in Fig.~\ref{fig:fig3} reveal the remarkable phenomenon that the topology of exceptional points associated
with the spectrum of the linearized equations induces coupling-driven bistability in the coupled oscillator system \cite{Perna2024-dg}. This bistability manifests as the coexistence of two distinct stable self-oscillating regimes. Moreover, encircling the EPs in parameter space allows for a controlled transition between these regimes.

\begin{figure}
    \centering
\includegraphics[width=8.5cm]{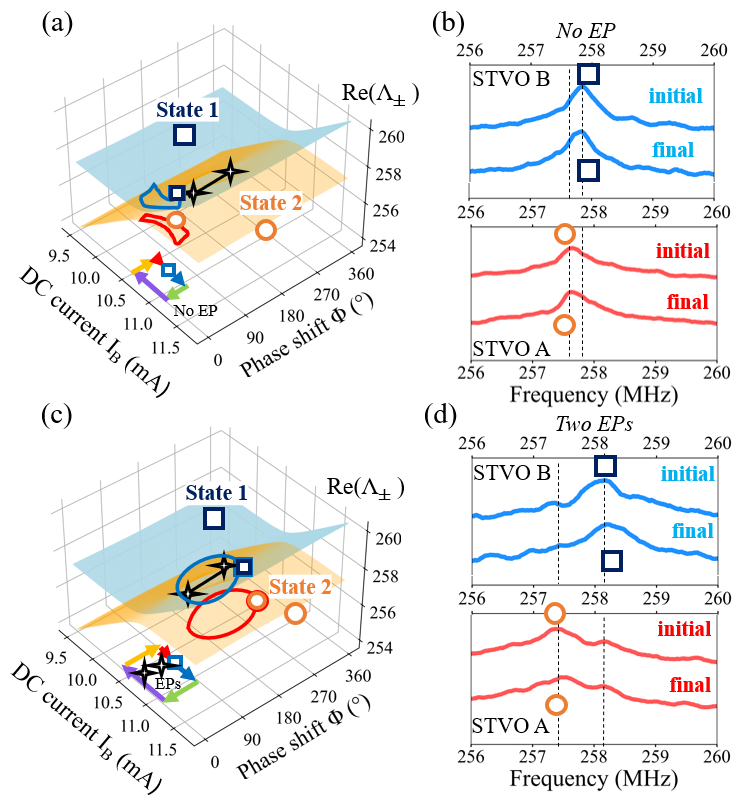}
\caption{\label{fig:fig4} Illustration of the absence of eigenstate exchange. Riemann sheet structure and corresponding trajectories not including an EP (a), and winding two EPs (c).  Closed loop without an EP (b), closed loops with 2 EPs (d). Describing a closed loop without an EP or with two EPs lead to no exchange of states. 
}
\end{figure}

In order to bring additional evidences that we successfully encircle a single EP, we perform two additional experiments presented in Fig.~\ref{fig:fig4} for which no exchange of eigenvalues is expected. As for the experiments presented in Fig.~\ref{fig:fig3}, the oscillators' RF output are recorded independently. In the first one (see Fig.~\ref{fig:fig4}(a-b)), we use the same experimental protocol as before, but this time for a 1-turn closed loop that does not include a single EP or branch cut, which means a smaller trajectory with a reduced $\delta \Phi$. In this case, as shown in Fig.~\ref{fig:fig4}(b), the final state is equivalent to the initial one for each oscillator.  For the second experiment (see Fig.~\ref{fig:fig4}(c-d)), we define a protocol that includes 1-turn around two EPs. In this latter case (see Fig.~\ref{fig:fig4}(d)), we also recover the initial state. Indeed, the higher frequency state remains in STVO $B$, and the lower frequency state stays in STVO $A$ after encircling two EPs instead of only one. 


In conclusion, we define an experimental protocol and a theoretical framework to identify and control the encirclement of exceptional points (EPs) in coupled vortex-based spintronic nano-oscillators.  Our successful experimental demonstration rely on the control of level repulsion and attraction between STVOs, respectively, associated with conservative and dissipative coupling. Moreover, we showcase at room temperature the encirclement of an EP in spintronic systems, where navigating around an EP leads to state switching—a promising mechanism for non-reciprocal transport applications \cite{doppler_dynamically_2016}. Investigating the coupling mechanism and exploring the topological nature of the exceptional point is essential for designing new detection-type devices with innovative uses of gain/loss systems, non-reciprocal transport, and chiral transfer \cite{ Ghosh2016-gx, yuan_periodic_2023}. Such properties can be explored in spintronic systems through spin current injection and control of non-conservative dynamics. Our work contributes to understanding non-Hermitian physics, such as mode selection or non-trivial topology, in a very tunable and simplified EP-based spintronic system at room temperature.

\smallbreak
KH acknowledges financial support from the ANRT contract number 2021/1341. This work has been supported by a France 2030 government grant managed by the French national research agency (ANR) PEPR SPIN "SPINCOM" (ANR-22-EXSP-0005). SW acknowledges financial support from the Helmholtz Young Investigator Group Program (VH-NG-1520).

\end{document}